\documentclass[twocolumn,showpacs,preprintnumbers,amsmath,amssymb]{revtex4}

\usepackage{graphicx}
\usepackage{dcolumn}
\usepackage{bm}

\begin{document}

\preprint{APS/123-QED}

\title{Lattice dynamics and electron-phonon coupling in transition metal diborides}
\author{R. Heid$^{1}$, B. Renker$^{1}$, H. Schober$^{2}$, P. Adelmann$^{1}$,
        D. Ernst$^{1}$, and K.-P. Bohnen,$^{1}$ }
\affiliation{$^{1}$Forschungszentrum Karlsruhe, IFP, P.O.B. 3640,
       D-76021 Karlsruhe, Germany}
\affiliation{$^{2}$Institut Laue-Langevin, BP 156 X, F-38042
       Grenoble Cedex, France }

\date{\today}

\begin{abstract}
The phonon density-of-states of transition metal diborides
TMB$_2$ with TM = Ti, V, Ta, Nb and Y has been measured using the
technique of inelastic neutron scattering. The experimental data
are compared with ab initio density functional calculations
whereby an excellent agreement is registered. The calculations
thus can be used to obtain electron-phonon spectral functions
within the isotropic limit. A comparison to similar data for
MgB$_2$ and AlB$_2$ which were subject of prior publications as
well as parameters important for the superconducting properties
are part of the discussion.

\end{abstract}

\maketitle

The discovery of superconductivity in MgB$_2$ at the unexpected
high temperature of 39\,K has renewed the interest in diboride
compounds. The well known AlB$_2$ structure ($P6/mmm$) is formed
with quite a number of transition metals (TM) \cite{Buzea}.
However, a clear proof of superconductivity could not be
established for any of these candidates. Obviously the influence
of lattice faults on the transport properties is important. A
unique feature for MgB$_2$ is the depopulation of B 2$p-\sigma$
bands due to a particular Mg-B interaction \cite{Kortus}. It has been shown by
band structure calculations for the TM-diborides that $E_F$ moves
up into the TM 4$d$-states with the consequence that these
compounds behave more like normal metals \cite{Vajeeston}. Most
important for the transport properties is the electron-phonon
interaction in these systems. Very recently results from point
contact spectroscopy have been reported for TM = Zr, Nb, and Ta.
The authors conclude on an only modest electron-phonon
interaction parameter for these samples \cite{Naidyuk}. A general
investigation of the phonon spectra of TM-diborides has so far
not be performed. In the present paper we confront measurements
of the generalized phonon density of states for selected
compounds with TM = Ta, V, Nb, Ti, and Y with ab initio density
functional theory (DFT) calculations. Results for electron-phonon
spectral functions are derived and placed in context to previous
investigations of superconducting MgB$_2$.

All diborides were prepared from stoichiometric mixtures of TM
elements and amorphous $^{11}$B. We have chosen the less
absorbing $^{11}$B in view of the intended inelastic neutron
scattering (INS) experiments. The mixed powders were compacted to
pellets and arc melted. The final polycrystalline samples showed
metallic brightness and were found to be single phase by $x$-ray
diffraction. All of our samples failed to show superconductivity
above 4\,K although transition temperatures for the pure elements
of Nb and Ta with 9.2\,K and 4.39\,K are known. Our INS
experiments were performed on the IN6 time-of-flight spectrometer
at the HFR in Grenoble, France, with an incident neutron energy
of 4.75\,meV at 300\,K in the upscattering mode. A high chopper
speed of 201\,rms and focusing in the inelastic region was used to
improve the resolution. The generalized phonon density of states
(GDOS) has been calculated from the recorded intensities
integrated over a scattering region from 14$^{\circ}$ to
114$^{\circ}$. For the data evaluation we have applied
multi-phonon corrections in a self-consistent procedure. The GDOS
implies  a weighting of vibrational modes of by $\sigma/m$
(scattering cross-section over the mass, see Tab.~\ref{tab1}).
For comparison the scattering power of $^{11}$B is
0.525\,barn/amu.

\begin{table}
   \caption{Structural and scattering parameters for selected
   diborides. Distances are given in \AA. The values in brackets apply
   to optimized geometries found in the DFT calculations after
   energy optimization.
   M-B and B-B denote the shortest metal-boron and boron-boron distances, respectively,
   and m is the TM mass number.
   $\sigma$/m is the neutron scattering cross-section over the mass in barn/amu.}
 \label{tab1}
  \begin{ruledtabular}
   \begin{tabular}{cccccccc}
  & MgB$_2$ & AlB$_2$ & TaB$_2$ & VB$_2$ & NbB$_2$ & TiB$_2$ & YB$_2$ \\
\hline
     a & 3.084 & 3.009 & 3.08  & 2.998 & 3.09  & 3.038 & 3.290 \\
       &(3.056)&(2.965)&(3.08) &(2.979)&(3.093)&(2.998)&(3.254)\\
     c & 3.522 & 3.262 & 3.265 & 3.056 & 3.3   & 3.23  & 3.835 \\
       &(3.622)&(3.232)&(3.272)&(2.995)&(3.337)&(3.188)&(3.830)\\
   M-B& 2.504 & 2.383 & 2.421 & 2.297 & 2.477 & 2.409 & 2.716 \\
   B-B & 1.781 & 1.737 & 1.799 & 1.772 & 1.794 & 1.773 & 1.914 \\
   m   &24.312 &26.982 &180.948&59.42  &92.906 &47.90  & 88.91  \\
   $\sigma$/m &0.151 &0.056& 0.033& 0.098& 0.067& 0.085& 0.087
    \end{tabular}
   \end{ruledtabular}
\end{table}

\begin{figure}[t]
\includegraphics[width=0.8\linewidth]{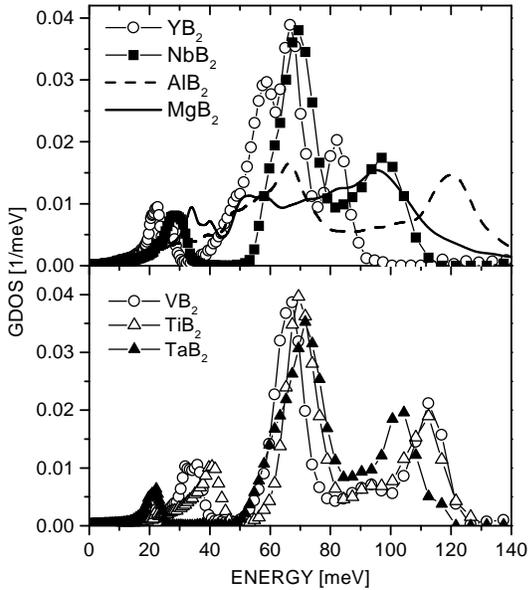}
\caption{The experimental generalized phonon density-of-states for
various transition metal diborides. AlB$_2$ and superconducting
MgB$_2$ are shown for comparison. Very significant changes are
related to structural differences. Signature of strong
electron-phonon coupling is the dramatic broadening of the
highest energy B peak visible for MgB$_2$ but not for the other
candidates.} \label{fig1.eps}
\end{figure}

\begin{figure}
\includegraphics[width=0.8\linewidth]{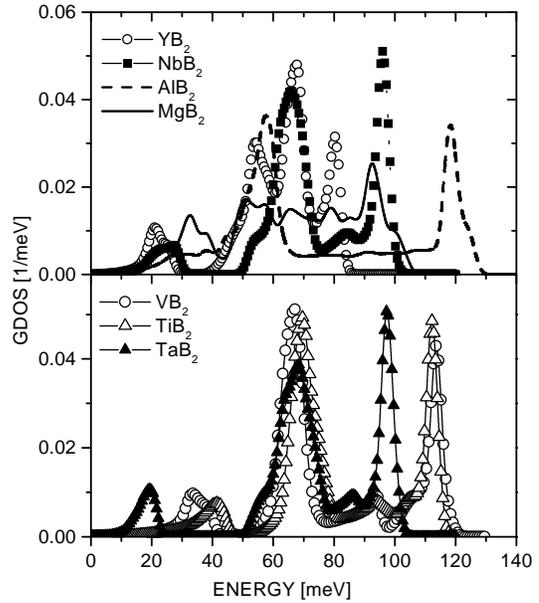}
\caption{Calculated GDOS spectra for all diborides shown in Fig.
1. The proper $\sigma$/m values have been applied for a direct
comparison to the experimental spectra.} \label{fig2.eps}
\end{figure}

Fig.\,1 shows a comparison of GDOS spectra for all investigated
TM-diborides. Each spectrum is normalized to 1.
The most significant common feature is a gap around 40 to 50\,meV.
When analyzing the respective intensities we find that the
high-frequency band is essentially composed of boron vibrations.
The gap is thus a consequence of the large mass difference, which
leads to a decoupling of transition metal and boron vibrations. A
clearly different behavior is observed for MgB$_2$ and AlB$_2$
where due to the smaller mass of the metal atoms a hybridization
of metal and boron modes occurs. In the low-frequency band of the
transition metal diboride spectra we observe distinct shifts
which however do not simply follow a $\sqrt{m}$ relation. This
indicates that there are appreciable differences in the B-TM
coupling. This conclusion is confirmed by the concomitant
variations of interatomic distances (Tab.~\ref{tab1}). Somewhat more
surprising is the fact that significant frequency shifts are
equally registered for the peak at the high-frequency end of the
spectra where the lowest frequency is found in YB$_2$. The
E$_{2g}$ and B$_{1g}$ modes, which are important for
superconductivity in MgB$_{2}$, lie within this region where
neighboring boron atoms move against each other. Strong
electron-phonon coupling is expected to cause a significant
broadening. This broadening is visible in the spectrum of MgB$_2$
but not in those of the other compounds. The low frequency of the
peak in YB$_{2}$ alone can, therefore, not be taken as a
signature of strong electron-phonon coupling. A simpler
explanation is offered by comparatively weak B-B-interactions. As
can be seen from Tab.~\ref{tab1} the B-B distance is exceptionally large
for this compound. If any, then the $G(\omega)$ for NbB$_2$ shows
signatures of a stronger electron-phonon coupling, a compound for
which some authors report superconductivity.

First principals density-functional calculations give parameter
free insight into the electronic structure of all of these
compounds. They in addition yield the lattice dynamical
properties. Corresponding results are shown in Fig.\,2. For the
calculations we used the mixed basis pseudopotential method,
which is described in some detail in Ref. \cite{Heid,Bohnen}.
Norm-conserving pseudopotentials for V, Ta, Nb, Y and Ti were constructed
according to the description of Bachelet-Hamann-Schl\"uter
\cite{Bachelet,Ho}, whereas for boron a Vanderbilt-type potential
was created \cite{Vanderbilt}. The mixed basis scheme uses a
combination of local functions together with plane waves for a
representation of the valence states. The good agreement between
calculated and experimental structural parameters  (see Tab.~\ref{tab1})
as well as between calculated and experimental phonon
GDOS (Figs.\,1 and 2) proves the validity of
our description.

A very unique feature of electron-phonon coupling in  MgB$_2$ is
the downshift of the in-plane E$_{2g}$ mode well below the
out-of-plane B$_{1g}$ mode. Such an inversion of the usual
sequence of mode frequencies is not encountered in any of the
other diborides as shown in Tab.~\ref{tab2}, where we compile the
calculated $\Gamma$-point frequencies of the boron vibrations.

\begin{table}
   \caption{Calculated $\Gamma$ point frequencies in meV.}
 \label{tab2}
  \begin{ruledtabular}
   \begin{tabular}{cccccccc}
  & MgB$_2$ & AlB$_2$ & TaB$_2$ & VB$_2$ & NbB$_2$ & TiB$_2$ & YB$_2$ \\
\hline
     E$_{1u}$ & 40.5 & 36.6 & 52.6 & 60.6  & 52.0 & 65.5 & 44.1 \\
     A$_{2u}$ & 50.2 & 52.1 & 61.9 & 62.1  & 60.5 & 66.4 & 45.3 \\
     E$_{2g}$ & 70.8 &125.0 &100.6 &114.9  & 98.4 &112.8 & 75.5 \\
     B$_{1g}$ & 87.0 & 61.3 & 68.8 & 69.6  & 69.8 & 70.0 & 75.1
       \end{tabular}
   \end{ruledtabular}
\end{table}

The present calculations prove capable of describing the lattice
dynamics of various diborides with quite different electronic
transport properties.
For the TM diborides, the isotropic character of the d-bands
has the consequence that these compounds behave electronically
very much like normal 3D metals \cite{Vajeeston}.
Although the band structures show strong
similarities, the Fermi level $E_F$ moves up and down within the
region of the transition metal $d$-bands. This variable band
filling leads to appreciable differences in the various
interatomic distances.
The E$_{2g}$ frequencies exhibit an expected inverse correlation with
the in-plane lattice constant, which indicates that the B-B covalent
bond is electronically rather similar for all these compounds.
On the contrast, the out-of-plane boron mode B$_{1g}$ is very insensitive
to changes in the lattice constants.

Somewhat
exceptional in our series of transition metal diborides is only
YB$_2$. Peculiarities in the electronic structure lead to a
drastic increase of both the a and c lattice constants.
As a consequence of the increased B-B bond length, the longitudinal
force constant of the B-B bond is strongly reduced, and becomes even
smaller than the transverse coupling.
This is the reason for the observed softening of the high-frequency
B-modes in YB$_2$.

\begin{figure}
\includegraphics[width=0.8\linewidth]{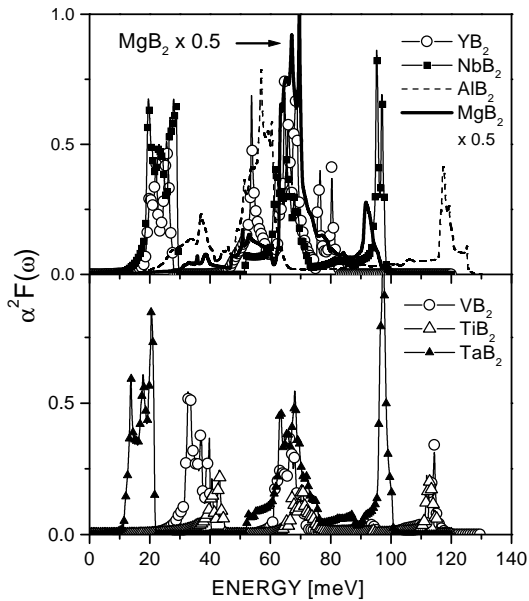}
\caption{Isotropic Eliashberg functions for the investigated
diborides as obtained in the present calculation.}
\label{fig3.eps}
\end{figure}

In order to get a more quantitative picture of the electron-phonon
coupling within the series of diborides we have calculated the
isotropic Eliashberg function $\alpha^{2}$F($\omega$) which
describes the phonon mediated pairing interaction (Fig.\,3).
Details of the computations are outlined in Ref.~\cite{Heid}.
Again MgB$_2$ proves to be singular. The extremely strong
coupling at high-phonon frequencies is known to be connected to
the E$_{2g}$ boron in-plane mode. It is found as a general trend
that the region of strong coupling for TM diborides shifts down
to lower energies as found for conventional superconductors thus
leaving hope only for modest T$_c$'s.
The existence of 2D- and 3D-type Fermi sheets derived mainly
from  boron $\sigma$ and $\pi$ bands, respectively,
has been shown to be important for a quantitative
understanding of the transport properties of MgB$_2$ \cite{Tsuda}.
The 2D Fermi surface is connected to the hole doping of boron $2p$ bands
and does not seem to be important for transition metal diborides.

\begin{table}
   \caption{Average coupling constant $\lambda$, effective average phonon
   frequency $\omega_{log}$ (in meV), and estimates of T$_c$ from the
   linearized gap equation.  N(0) denotes the density of states at the Fermi energy
   (per unit cell and spin).}
 \label{tab3}
  \begin{ruledtabular}
   \begin{tabular}{ccccc}
  & N(0) & $\lambda$ & $\omega$$_{log}$ &  T$_c$ ($\mu$*=0.13)  \\
\hline
    MgB$_2$ & 0.335  & 0.73  & 60.9 &  21.7  \\
    AlB$_2$ & 0.184  & 0.43  & 49.9 &  2.3   \\
    TaB$_2$ & 0.452  & 0.79  & 25.8 &  10.6  \\
     VB$_2$ & 0.592  & 0.28  & 44.1 &  $<$ 1   \\
    NbB$_2$ & 0.520  & 0.67  & 30.5 &  8.4   \\
    TiB$_2$ & 0.179  & 0.10  & 52.9 &  -  \\
     YB$_2$ & 0.560  & 0.46  & 37.4 &  2.4\\
   \end{tabular}
   \end{ruledtabular}
\end{table}

We have also calculated the relevant
parameters for the description of the superconducting properties
within the Eliashberg formalism. The results are compiled in
Tab.~\ref{tab3}.
A common feature unfavorable for superconductivity found
in all transition metal diborides is the comparatively low
$\omega_{log}$. This quantity  represents the effective average
frequency of the coupling modes and sets the energy scale for the
pairing interaction.
Its small value as compared to MgB$_2$ indicates that the pairing
interaction is mainly mediated by the TM vibrations and not by the
boron modes.
The isotropic coupling constant $\lambda$ shows a variation from weak
to intermediate coupling strength.
In Tab.~\ref{tab3}, we have also included values for the superconducting
transition temperature as obtained by solving the linearized isotropic
gap equations.
The T$_c$ values depend strongly on the screening properties expressed
by $\mu^{*}$ and drop appreciably for $\mu^{*} > 0$.
The values shown for a typical metallic screening of $\mu^{*}=$0.13 thus
should only be taken as indicators of the general trend.
Furthermore, as mentioned before, for MgB$_2$ it is necessary to
go beyond the isotropic limit and take into account the multigap
structure introduced by the particular Fermi surface geometry for a
proper quantitative description of the pairing state.
Anisotropy is expected to play a minor role for
the coupling in the low-frequency region of the transition metal
vibrations that are dominated by the 3D-metal bands.

The values of $\lambda$ and $\omega_{log}$ are most
favorable for MgB$_2$.
The next promising candidates according to
the present calculations are TaB$_2$  and NbB$_2$, for which some
authors find superconductivity at 9.5\,K and 6.4\,K,  respectively \cite{Heid}.
Differences in stoichiometry might be able to explain different
experimental results. As far as a comparison to recent point-contact
measurements is possible we can state that the maxima in our $G(\omega)$
correspond well to the peaks found in the second derivative of the $I-V$
characteristic.
There is also agreement in a strongly reduced
electron-phonon coupling strength for transition metal diborides
with respect to MgB$_{2}$. We, however, want to mention that the
reported very small numerical values for $\lambda_{PC}$ should not
be compared directly to our values since point contact
experiments do not measure the McMillan $\lambda$.

To summarize, we have presented measurements of the generalized phonon
density-of-states for various transition metal diborides.
The observed trends in the lattice dynamics can be well ascribed by
first principles calculations and do not exhibit indications of strong
electron-phonon interaction.
This view is supported by theoretical calculations of the electron-phonon
coupling.
While they do not exclude the possibility of superconductivity with
low T$_c$ mediated by TM vibrations, they underline the outstanding
properties of MgB$_2$ among the class of diborides.

\end{document}